# BCB Based Packaging for Low Actuation Voltage RF MEMS Devices

*David Peyrou, Fabienne Pennec, Hikmat Achkar, Patrick Pons, Fabio Coccetti, Hervé Aubert, Robert Plana*

## ABSTRACT

This paper outlines the issues related to RF MEMS packaging and low actuation voltage. An original approach is presented concerning the modeling of capacitive contacts using multi-physics simulation and advanced characterization. A similar approach is used concerning packaging development where multi-physics simulations are used to optimize the process. A devoted package architecture is proposed featuring very low loss at microwave range.

## INTRODUCTION

RF MEMS have demonstrated, during the last ten years, very attractive potential to allow the introduction of "intelligence" in the RF front end architecture [1] through analog signal processing techniques. Nevertheless, those devices with moveable structures still have some issues to be successfully introduced at the industrial level. The first issue deals with the actuation medium and the corresponding reliability. Today, it is well known that membranes and cantilevers can be actuated through electrostatic, thermal, magnetic and piezoelectric forces. Each of these types of actuators features benefits and drawback and it seems that today electrostatic actuators could offer the best trade-off when issues concerning the very high actuation voltage and reliability issues are solved [2-5]. It has already been demonstrated that packaging improves the life time of RF MEMS and it is important to develop a process that is fully compatible with MEMS process and that introduces very low insertion loss. This paper will outline the issues that are important to assess a packaged low actuation voltage RF MEMS.

## LOW ACTUATION VOLTAGE ISSUE

This section will address issues related to low actuation voltage RF MEMS. A capacitive switch is given as an example as shown in figure 1. The upper moveable membrane locally modifies the RF impedance when an electrostatic force is applied. The performance of this type of component is driven by the quality of the contrast between the on state (very low capacitance) and the off state (very high capacitance) and by the reproducibility of the capacitive contact. The quality is reflected by the ratio between the theoretical capacitance (i.e MIM) and the obtained capacitance.

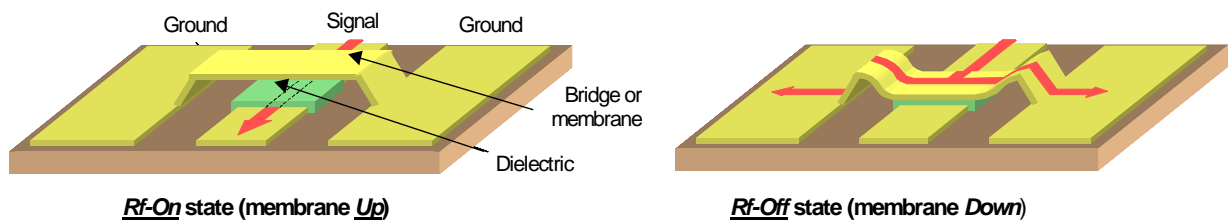

Figure 1 : Topologies of a capacitive switch

The actuation voltage can be expressed as the following :

$$Vp = \sqrt{\frac{8k}{27\varepsilon_0 A} go^3} \quad (1)$$

Where k represents the stiffness of the membrane, go the height and A the actuation surface area. It is clear that lower actuation voltage could be obtained through low height, low stiffness and large actuation area. Lowering the membrane height is not a good option as it impacts directly on the microwave performance of the device. Extending the actuation area is feasible but again it translates to a device featuring larger area that can influence the microwave behaviour, we can see that the only parameter that can be tuned is the stiffness of the upper membrane. This can be obtained through the use of materials featuring low young's modulus or through appropriate design of the actuator and the membrane.

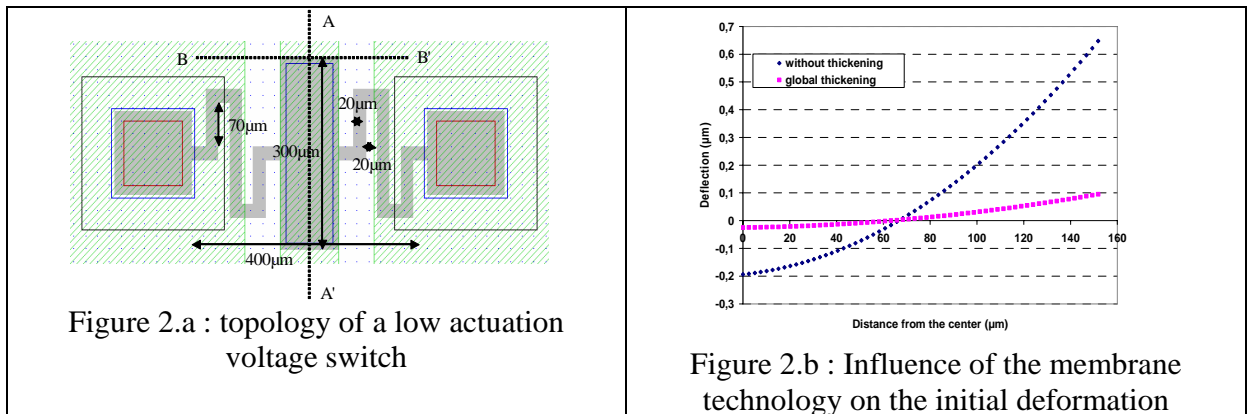

Figure 2.a : topology of a low actuation voltage switch

Figure 2.b : Influence of the membrane technology on the initial deformation

Figure 2.a shows a topology of capacitive switch that has been designed and fabricated. The device has a gold membrane composed of 50nm of evaporated gold and 1.5 µm of electroplated gold and a Si3N4 dielectric layer to form the capacitive contact. Special attention has been paid to monitor and to control the overall stress in the membrane. Evaporated gold exhibits a compressive stress of -75MPa and a tensile stress of 7.5 MPa is reported for the electroplated gold. The structures fabricated exhibit an actuation voltage ranging from 5 to 10 volts. It has been observed that due to residual stress, a significant initial deformation has been obtained during the release of the membrane as reported in figure 2.b. In order to overcome this problem, the metallization was thickened from 1.5 to 4µm which resulted in a very flat membrane.

The other issue related to the low actuation voltage deals with the contact quality which results in a lower capacitance than expected due to flatness and/or roughness problems.

In order to illustrate this behavior, we have reported roughness measurements done on different tests vehicles in figure 3. From these measurements, an effective air gap ranging from 10 nm to 30 nm can be calculated, which results in a serial capacitance that explains the difference between the theoretical and measured values.

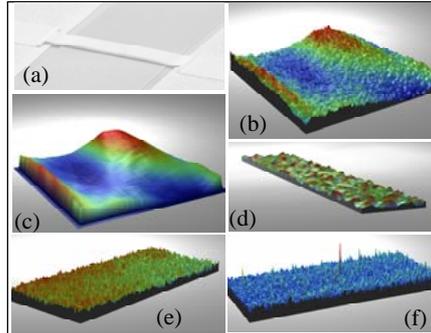

Figure 3. Roughness characterization for RF MEMS switch – (a) RF MEMS switch, bottom surface of the beam (b) Profile without filter and (c) Profile with low pass filter, dielectric surface deposited on three different gold layer as central conductor (d) Evaporated gold 2000 Å, (e) Evaporated gold 2 µm and (f) Electrolytic gold 2.5 µm

We developed a reverse engineering method to discover the origin of the mismatch between the theoretical down state capacitance and measurements. We used an optical profilometer (VEECO) to capture three-dimensional (3D) data points of the dielectric surface. Then, using Matlab and COMSOL functions, we convert the surface data into stereolithography format which is an ASCII standard format used in CAD software.

Figure 4 illustrates the full method and data flow : dielectric roughness is scanned by the optical profilometer (A) to generate an ASCII file (B) representative of 3D coordinates points of the surface. Than solid model (C) and object file (E) as STL, VRML or Comsol geometry can be obtained using Matlab. To enhance the shape, one can use (D) some filters (filter2 for MatLab or a CATIA CAD Software).

After this, the model is ready to be imported into a finite element software (F). We used multi-physics software provided by COMSOL, to set up the electrostatic model (materials and boundaries conditions (F) - meshing (G), results and post processing (H)).

Using this method, we have found an accurate agreement between roughness measurements and multi-physics simulations. This method will be used to optimize capacitive contacts and ohmic contacts in the future.

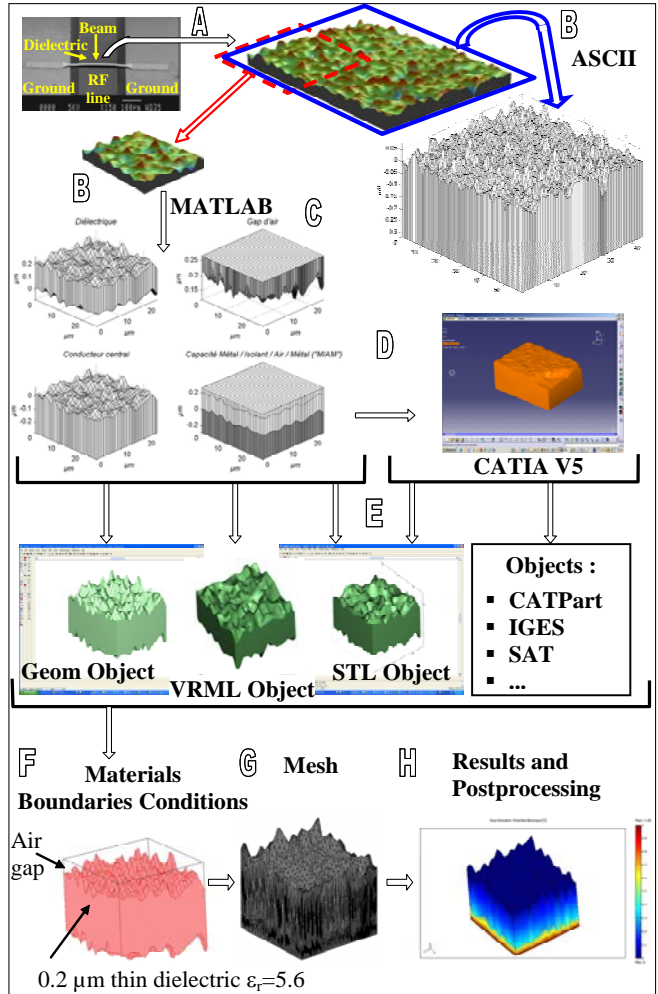

Figure 4 : Description of the reverse engineering method

The next section of the paper deals with the packaging approach (technology and design) that has been developed to package the RF MEMS devices

**PACKAGING APPROACH**

As previously stated, the packaging approach has been driven by requirements for compatibility with the existing RF MEMS processes and by the need for the process to package multi-scale devices with very low insertion loss in the microwave range. Assuming these requirements, we have chosen a BCB based package as it is a very flexible process and features very attractive microwave properties. Figures 5 and 6 summarize the topologies of the package. The package technology is based on BCB sealing ring associated with a Foturan cap. Both the BCB ring and the Foturan package must have appropriate dimensions to be relevant with the RF requirements.

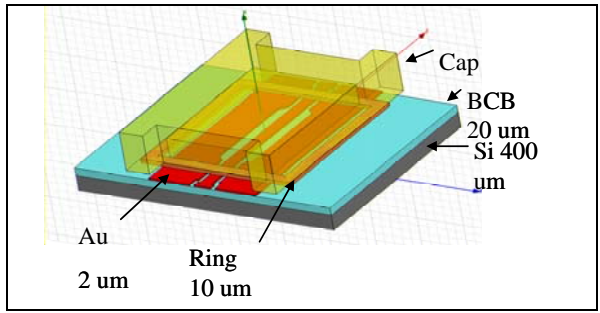

Figure 5 : BCB based packaging topology

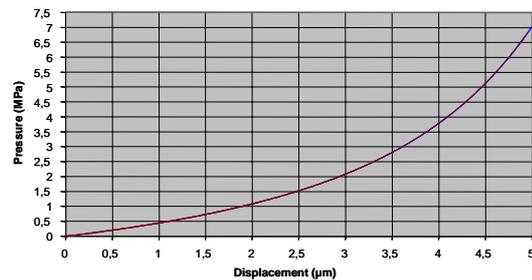

Figure 6 : Foturan package

The BCB ring has to ensure some hermeticity; this requirement will influence the thickness of the BCB ring. It will also influence the assembling process that has to apply enough force to fill the coplanar wave guide gap to feature a homogenous layer and to not exceed some pressure for reliability issues.

A model has been developed to simulate the assembly process and to calculate the pressure that is necessary to fill the CPW gap with the BCB ring. Figure 7 illustrates the model that has been set up and figure 8 presents simulations of the pressure versus displacement. In our case, we have a metallization featuring 2.3µm thickness that corresponds to a pressure in the 1.2 MPa range to fill the gap correctly.

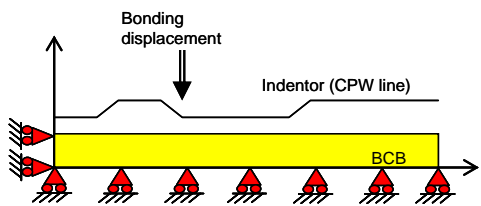

Figure 7 : model developed to simulate the gap filling with BCB

Figure 8 : simulation of the pressure versus BCB displacement (µm)

Electromagnetic simulations have been carried out using HFSS in order to determine the most appropriate architecture for the package. Figure 7 shows the return loss magnitude versus frequency for different wall thicknesses. The results show that wall thicknesses up to 200µm are compatible with microwave requirements. Figure 8 displays the frequency evolution of the insertion losses for different BCB ring height and the results indicate that BCB rings ranging from 5 to 20 µm are compatible with RF and microwave requirements.

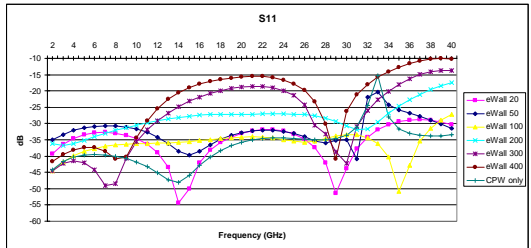
Figure 7 : return loss evolution versus frequency for different wall thicknesses

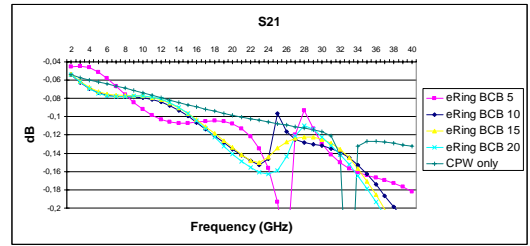
Figure 8 : Insertion loss evolution versus frequency for different BCB ring height

The package has been fabricated and has been assembled with a coplanar wave guide and the measurements show very good agreement with respect to the simulations. The results also show that the BCB ring and Foturan package are a very attractive solution for low cost and low insertion package. The other advantage of this packaging approach deals with the very low stress associated with its fabrication process and assembling that make it usable for low actuation RF MEMS.

**CONCLUSIONS**

This paper outlines both the important issues for lowering the actuation voltage and the low stress packaging to improve the reliability of RF MEMS devices. It has been shown that monitoring the stress was a very important issue to lower the actuation voltage, and a novel design has been demonstrated based on this approach. We also demonstrated that the contact quality was another issue and we proposed an original method based on surface characterization and multi-physics modelling that allows the optimization of the contact quality. Finally, we have investigated the issues for low stress packaging and we have proposed a BCB based packaging featuring low insertion loss in the microwave range.